\documentclass[conference]{IEEEtran}
\IEEEoverridecommandlockouts
\usepackage[table]{xcolor}
\usepackage{cite}
\usepackage{graphicx}
\usepackage{lipsum}
\usepackage{stfloats}
\usepackage{bm}
\usepackage{amsmath,amssymb,amsfonts}
\usepackage{amsthm}
\usepackage{pifont}
\usepackage{bbding}
\usepackage{booktabs}
\usepackage{textcomp}
\usepackage{ctable}
\usepackage{threeparttable}
\usepackage{footmisc}
\usepackage{framed} 
\usepackage{array}
\usepackage{multirow}
\usepackage{longtable}
\usepackage{rotating}
\usepackage{fancyhdr}
\usepackage{lastpage}
\usepackage{layout}
\usepackage{wrapfig}
\usepackage{xcolor}
\usepackage{tabularx}
\usepackage{tabu}
\usepackage{enumerate}
\usepackage{enumitem}
\usepackage{url}
\usepackage{dsfont}
\usepackage{algorithm}
\usepackage{algorithmic}
 \usepackage{enumerate}
 \usepackage{makecell}

\makeatletter
\newcommand{\removelatexerror}{\let\@latex@error\@gobble}
\makeatother

\graphicspath{{Figures/}}
\DeclareGraphicsExtensions{.png, .eps}

\def\BibTeX{{\rm B\kern-.05em{\sc i\kern-.025em b}\kern-.08em
    T\kern-.1667em\lower.7ex\hbox{E}\kern-.125emX}}

\begin{document}
	\title{Knowledge Base Aware Semantic Communication in Vehicular Networks}
	\author{\IEEEauthorblockN{Le Xia\IEEEauthorrefmark{1},
				 Yao Sun\IEEEauthorrefmark{1},
				 Dusit Niyato\IEEEauthorrefmark{2},
				 Kairong Ma\IEEEauthorrefmark{1},
				 Jiawen Kang\IEEEauthorrefmark{3},
				 and Muhammad Ali Imran\IEEEauthorrefmark{1}}
	\IEEEauthorblockA{\IEEEauthorrefmark{1}James Watt School of Engineering, University of Glasgow, Glasgow, UK\\
	\IEEEauthorrefmark{2}School of Computer Science and Engineering, Nanyang Technological University, Singapore\\
	\IEEEauthorrefmark{3}School of Automation, Guangdong University of Technology, Guangdong, China\\
	Email: Yao.Sun@glasgow.ac.uk}}

	\maketitle
	\begin{abstract}
	Semantic communication (SemCom) has recently been considered a promising solution for the inevitable crisis of scarce communication resources.
	This trend stimulates us to explore the potential of applying SemCom to vehicular networks, which normally consume a tremendous amount of resources to achieve stringent requirements on high reliability and low latency.
	Unfortunately, the unique background knowledge matching mechanism in SemCom makes it challenging to realize efficient vehicle-to-vehicle service provisioning for multiple users at the same time.
	To this end, this paper identifies and jointly addresses two fundamental problems of knowledge base construction (KBC) and vehicle service pairing (VSP) inherently existing in SemCom-enabled vehicular networks.
	Concretely, we first derive the knowledge matching based queuing latency specific for semantic data packets, and then formulate a latency-minimization problem subject to several KBC and VSP related reliability constraints.
	Afterward, a SemCom-empowered Service Supplying Solution (S$^{\text{4}}$) is proposed along with the theoretical analysis of its optimality guarantee.
	Simulation results demonstrate the superiority of S$^{\text{4}}$ in terms of average queuing latency, semantic data packet throughput, and user knowledge preference satisfaction compared with two different benchmarks.
	\end{abstract}

	\IEEEpeerreviewmaketitle
	
	\section{Introduction}
	Semantic communication (SemCom) beyond the conventional Shannon paradigm has recently been recognized as a promising remedy for communication resource saving and transmission reliability promotion~\cite{weaver1953recent,bao2011towards,xia2022wiservr,strinati20216g,xia2022wireless,9797984}.
	Such a trend, therefore, inspires us to investigate the potential of exploiting SemCom in vehicular networks to deal with the inevitable scarcity of available communication resources while enhancing information interaction efficiency, especially when considering large-scale vehicle-to-vehicle (V2V) communications.
	
	Different from bit-oriented traditional communication, the accurate delivery of semantics becomes the cornerstone of SemCom~\cite{weaver1953recent}.
	Taking a single SemCom-enabled V2V link as an example, a sender vehicle user (VUE) first leverages background knowledge relevant to source messages to filter out irrelevant content and extract core semantic features that only require fewer bits for transmission~\cite{bao2011towards}.
	Once the receiver VUE has the same knowledge as the sender, its local semantic interpreters are capable of accurately restoring the original meanings from the received bits, even with intolerable bit errors in data dissemination~\cite{xia2022wiservr}.
	Consequently, SemCom can significantly alleviate the resource scarcity problem, while ensuring sufficient transmission efficiency along with ultra-low semantic errors for each V2V transmission.
	Apart from these advancements, it should be noted that the background knowledge equivalence is of paramount importance to eliminate semantic ambiguity, which has led to a key concept of~\textit{knowledge base} (KB) in the realm of SemCom~\cite{strinati20216g,xia2022wireless,9797984}.
	Specifically, a single KB is deemed a small information entity that contains background knowledge corresponding to only one particular application domain (e.g., music or sports, etc.)~\cite{strinati20216g}.
	Hence, holding some common KBs becomes the necessary condition to perform SemCom between two VUEs in accordance with the knowledge equivalence principle.
 
	Recently, there have been some technical works on SemCom from the link level~\cite{xia2022wiservr} to the networking level~\cite{xia2022wireless} under cellular network architectures. 
	Nevertheless, to the best of our knowledge, no research pertinent to service provisioning has been conducted in~\textit{SemCom-enabled vehicular networks} (SCVNs), which should be rather challenging and involves two unique yet fundamental problems.
	To be concrete, due to the varying practical KB sizes, personal KB preferences, and limited vehicular storage capacities, the first problem is how to devise an optimal~\textit{knowledge base construction} (KBC) policy not only proactively but also collaboratively for all VUEs to construct their respective appropriate KBs for better service provisioning.
	In the meantime, when considering different types of KBs equipped on numerous VUEs and unstable wireless link quality, it can be challenging to well solve the service provisioning-driven VUE pairing problem to meet the knowledge matching restriction, thus shaping the second problem namely~\textit{vehicle service pairing} (VSP).
	
	To that end, in this paper, we propose a novel SemCom-empowered Service Supplying Solution (S$^{\text{4}}$) in SCVN.
	First, we theoretically derive the KB matching based queuing latency for semantic data packets received at each possible VUE pair.
	Afterward, a joint latency-minimization problem is mathematically formulated subject to several KBC and VSP-related reliability constraints and other practical system limitations.
	To address this problem with low complexity, we then develop an efficient solution S$^{\text{4}}$ and theoretically prove its optimality.
	Particularly, a primal-dual problem transformation method is exploited in S$^{\text{4}}$, followed by a two-stage method dedicated to solving multiple subproblems.
	Given the dual variable in each iteration, the first stage is to obtain the optimal KBC sub-policy for each potential VUE pair, whereby the second stage is to finalize the optimal solutions of KBC and VSP for all VUEs.
	Finally, numerical results demonstrate the performance superiority of S$^{\text{4}}$ in terms of average queuing latency, semantic data packet throughput, and user knowledge preference satisfaction compared with two different benchmarks.
    
    \addtolength{\topmargin}{0.02in}
    
	\section{System Model and Problem Formulation}
\begin{figure}[t]
		\centering
		\includegraphics[width=0.489\textwidth]{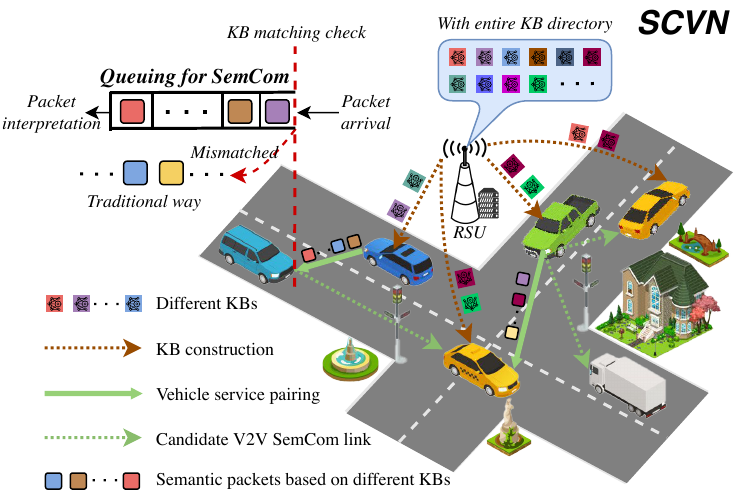} 
		\caption{The SCVN scenario and the knowledge base aware queuing model for semantic data packets transmitted between VUEs.}
		\label{SCVN}
    \end{figure}

	Considering an SCVN scenario as shown in Fig.~\ref{SCVN}, the total of $V$ VUEs are randomly distributed within the coverage of a single roadside unit (RSU), and each VUE $i \in \mathcal{V}=\left\{1, 2,\ldots,V\right\}$ is capable of providing SemCom-empowered services to others.
	Let $\gamma_{i,j}$ denote the signal-to-interference-plus-noise ratio (SINR) experienced by the V2V link between VUE $i$ and VUE $j$.
	With this, the set of communication neighbors of VUE $i$ is defined as $\mathcal{V}_{i}=\left\{j|j \in \mathcal{V},j\neq i, \gamma_{i,j} \geqslant \gamma_{0}\right\}$, $ i \in \mathcal{V}$, where $\gamma_{0}$ is a prescribed SINR threshold.
	
	\subsection{Semantic Service Provisioning Model}
	Due to the unique mechanism of semantic interpretation, the acquisition of necessary background knowledge is inevitable for all SemCom-enabled transceivers.
	In this work, assuming that each VUE $i$ proactively downloads and constructs its required KBs from the RSU under its finite KB storage capacity $C_{i}$.
	Meanwhile, suppose that there is a KB library $\mathcal{K}$ with a total of $N$ differing KBs in the considered SCVN, and each requires a unique storage size $s_{n}$, $n \in \mathcal{K}=\left\{1, 2,\ldots, N\right\}$.
	Furthermore, we define a binary KBC indicator as
	\begin{equation}
		\label{alpha}
			\alpha_{i}^{n}=\left\{\begin{aligned}
			1,\quad &  \text{if KB $n$ is constructed at VUE $i$;}\\
			0,\quad &  \text{otherwise.}
		\end{aligned}
		\right.
	\end{equation}
	It is worth mentioning that the same KB cannot be constructed repeatedly at one VUE to promote the storage efficiency.
	
	Besides, note that different VUEs may have different preferences for these KBs corresponding to their desired services.
	Without loss of generality, we assume that the KB popularity at each VUE meets Zipf distribution~\cite{piantadosi2014zipf}.\footnote{Other distributions with known probabilities can also be adopted without changing the remaining modeling and solution.}
	Hence, the probability of VUE $i$ requesting its desired KB $n$-based services (generating the corresponding semantic data packets) is $p_{i}^{n}=\left(r_{i}^{n}\right)^{-\xi_{i}}/\sum_{e \in \mathcal{K}}e^{-\xi_{i}}, \forall (i,n) \in \mathcal{V}\times \mathcal{K}$, where $\xi_{i}$ ($\xi_{i}\geqslant 0$) is the skewness of the Zipf distribution, and $r_{i}^{n}$ is the popularity rank of KB $n$ at VUE $i$.\footnote{The KB popularity rank of each VUE can be estimated based on its historical messaging records, which will not be discussed in this paper.}
	Based on $p_{i}^{n}$, we specially develop a KBC-related metric $\eta_{i}$ below, namely~\textit{knowledge preference satisfaction}, to measure the satisfaction degree of VUE $i$ constructing its interested KBs as
	\begin{equation}
		\eta_{i}=\sum_{n \in \mathcal{K}}\alpha_{i}^{n}p_{i}^{n}.
	\end{equation}
	It is further required that $\eta_{i} \geqslant \eta_{0}$, where $\eta_{0}$ is the unified minimum threshold that needs to be achieved at each VUE.
	
	Moreover, considering that each VUE can be paired with only one (another) VUE at a time.
	Let $\beta_{i\looparrowright j}$ denote the binary VSP indicator for a VUE $i$-VUE $j$ pair (suppose that VUE $i$ is the sender and VUE $j$ is the receiver), where
	\begin{equation}
		\label{beta}
			\beta_{i\looparrowright j}=\left\{\begin{aligned}
			1,\quad &  \text{if VUE $i$ is associated with VUE $j$;}\\
			0,\quad &  \text{otherwise.}
		\end{aligned}
		\right.
	\end{equation}
	Here, we use the notation $\looparrowright$ as an auxiliary illustration to specify the roles of sender and receiver in each VUE pair.
	
	\subsection{Knowledge Base Aware Queuing Model}
	As depicted in Fig.~\ref{SCVN}, the knowledge matching based semantic packet queuing delay is employed as the latency metric of SemCom, to characterize the average sojourn time of semantic data packets in the receiver VUE's queue buffer (following a first-come first-serve rule).
	Note that not all semantic packets arriving at the receiver VUE are always allowed to enter its queue, as some of them may mismatch the KBs currently held, rendering these packets uninterpretable.
	To avoid pointless queuing, these mismatched packets may have to choose the traditional communication way for information interaction, which will not be discussed in depth here.
	
	To preserve generality, we first suppose a Poisson data arrival process with average rate $\lambda_{i}^{n}=\lambda_{i}p_{i}^{n}$ for a sender VUE $i$ to account for its local semantic packet generation based on KB $n$, where $\lambda_{i}$ is the total arrival rate of all semantic packets at VUE $i$.
	Then we can obtain the effective packet arrival rate from sender VUE $i$ to receiver VUE $j$ as $\lambda_{i\looparrowright j}^{\mathit{eff}}=\sum_{n \in \mathcal{K}}\alpha_{i}^{n}\alpha_{j}^{n}\lambda_{i}^{n}$ by checking the KB matching status for all received semantic packets.
	Herein, denoting the ratio of the number of mismatched packets to the number of received packets in total for VUE $i$-VUE $j$ pair as $\theta_{i\looparrowright j}$, namely~\textit{knowledge mismatch degree}, which is explicitly calculated by
	\begin{equation}
		\label{kmd}
		\begin{aligned}
			\theta_{i\looparrowright j}&=\frac{\sum_{n \in \mathcal{K}}\alpha_{i}^{n}\left(1-\alpha_{j}^{n}\right)\lambda_{i}^{n}}{\sum_{n \in \mathcal{K}}\alpha_{i}^{n}\lambda_{i}^{n}}.
		\end{aligned}
	\end{equation}
		
	In parallel, let a random variable $I_{j}^{n}$ denote the Markovian interpretation time~\cite{lavee2009understanding} required by KB $n$-based packets at VUE $j$ with mean $1/\mu_{j}^{n}$, which is determined by the computing capability of the vehicle and the type of the desired KB.
	However, since multiple packets based on different KBs are allowed to queue at the same time, the interpretation time distribution for a receiver VUE should be treated as a general distribution.
	If further taking into account the KB popularity, we can calculate the ratio of the amount of KB $n$-based packets to the total packets in the VUE $i$-VUE $j$ pair's queue by $\epsilon_{i\looparrowright j}^{n}=p_{i}^{n}/\sum_{f \in \mathcal{K}}\alpha_{i}^{f}\alpha_{j}^{f}p_{i}^{f}$.
	Keeping in mind the independence among packets based on different KBs, the interpretation time required by each packet in the queue is now phrased as $W_{i\looparrowright j}=\sum_{n \in \mathcal{K}}\alpha_{i}^{n}\alpha_{j}^{n}\epsilon_{i\looparrowright j}^{n}I_{j}^{n}$.
	
	In full view of the above, the queue of each VUE pair in the SCVN fully qualifies as an M/G/1 model.
	According to the~\textit{Pollaczek-Khintchine formula}~\cite{ross2014introduction}, the average queuing latency for VUE $i$-VUE $j$ pair (denoted as $\delta_{i\looparrowright j}$) should be\footnote{In order to guarantee the steady-state of the queuing system, a condition of $\lambda_{i\looparrowright j}^{\mathit{eff}}\mathds{E}\left[W_{i\looparrowright j}\right]<1$ must be reached before proceeding. In this work, we assume that the packet interpretation rate is larger than the packet arrival rate to make the queuing latency finite and thus solvable.}
	\begin{equation}
		\delta_{i\looparrowright j}=\frac{\lambda_{i\looparrowright j}^{\mathit{eff}}\cdot \left(\mathds{E}^{2}\left[W_{i\looparrowright j}\right]+\mathit{Var}\left(W_{i\looparrowright j}\right)\right)}{2\left(1-\lambda_{i\looparrowright j}^{\mathit{eff}}\cdot \mathds{E}\left[W_{i\looparrowright j}\right]\right)}.\label{PK}
	\end{equation}
	On this basis, again leveraging the independence of $I_{j}^{n}$ over $n$, we can obtain the expectation of the interpretation time for all semantic data packets in the queue by
	\begin{equation}
		\mathds{E}\left[W_{i\looparrowright j}\right]=\sum_{n \in \mathcal{K}}\frac{\alpha_{i}^{n}\alpha_{j}^{n}\epsilon_{i\looparrowright j}^{n}}{\mu_{j}^{n}},\label{PK1}
	\end{equation}
	and the variance of $W_{i\looparrowright j}$ is given by
	\begin{equation}
		\begin{aligned}
		\mathit{Var}\left(W_{i\looparrowright j}\right)=\sum_{n \in \mathcal{K}}\alpha_{i}^{n}\alpha_{j}^{n}\left(\frac{\epsilon_{i\looparrowright j}^{n}}{\mu_{j}^{n}}\right)^{2}.\label{PK2}
		\end{aligned}
	\end{equation}
	Based on~(\ref{PK1}) and (\ref{PK2}), $\delta_{i\looparrowright j}$ in~(\ref{PK}) can be calculated explicitly.
	
	\subsection{Problem Formulation}
	According to the above discussions, we identify and formulate below a joint queuing latency-minimization problem under the KBC indicator set $\bm{\alpha}=\left\{\alpha_{i}^{n}|i \in \mathcal{V},n \in \mathcal{K}\right\}$ and the VSP indicator set $\bm{\beta}=\left\{\beta_{i\looparrowright j}|i \in \mathcal{V},j \in \mathcal{V}_{i}\right\}$.
    \begin{align}
	\mathbf{P0}:\ \min_{\bm{\alpha},\bm{\beta}} \quad & \sum_{i \in \mathcal{V}}\sum_{j \in \mathcal{V}_{i}}\beta_{i\looparrowright j}\delta_{i\looparrowright j} ~\label{P0}\\
	{\rm s.t.} \quad & \sum_{n \in \mathcal{K}} \alpha_{i}^{n}\cdot s_{n}\leqslant C_{i},\  \forall i\in \mathcal{V},\tag{\ref{P0}a}\\
	& \eta_{i}\geqslant \eta_{0},\ \forall i \in \mathcal{V},\tag{\ref{P0}b}\\
	&\sum_{j \in \mathcal{V}_{i}}\beta_{i\looparrowright j}= 1,\ \forall i \in \mathcal{V},\tag{\ref{P0}c}\\
	& \beta_{i\looparrowright j}=\beta_{j\looparrowright i},\ \forall \left( i,j\right) \in \mathcal{V}\times \mathcal{V}_{i},\tag{\ref{P0}d}\\
	& \sum_{j \in \mathcal{V}_{i}}\beta_{i\looparrowright j}\theta_{i\looparrowright j}\leqslant\theta_{0},\ \forall i \in \mathcal{V},\tag{\ref{P0}e}\\
	& \alpha_{i}^{n}\in \left\{ 0,1\right\},\ \forall \left( i,n\right) \in \mathcal{V}\times \mathcal{K},\tag{\ref{P0}f}\\
	& \beta_{i\looparrowright j}\in \left\{ 0,1\right\},\ \forall \left( i,j\right) \in \mathcal{V}\times \mathcal{V}_{i}.\tag{\ref{P0}g}
	\end{align}
	Constraint (\ref{P0}a) indicates the storage limitation of VUEs, while constraint (\ref{P0}b) corresponds to the aforementioned knowledge preference satisfaction requirement for each VUE.
	Constraints (\ref{P0}c) and (\ref{P0}d) mathematically model the single-association requirement of VUEs.
	Constraint (\ref{P0}e) represents that the knowledge mismatch degree of each VUE pair should be over the threshold $\theta_{0}$.
    Constraints (\ref{P0}f) and (\ref{P0}g) characterize the binary properties of $\bm{\alpha}$ and $\bm{\beta}$, respectively.
    As such, the main difficulty in solving $\mathbf{P0}$ lies on its combinatorial nature and non-convexity of its highly complicated objective function.
  	
  	\section{Proposed SemCom-Empowered Service Supplying Solution}
  	In this section, we illustrate how to design our proposed solution S$^{\text{4}}$ to efficiently cope with $\mathbf{P0}$ for SCVNs.
 	First, a Lagrange dual method is leveraged to eliminate the cross-term constraints in $\mathbf{P0}$ with a corresponding dual optimization problem transformed (referring to $\mathbf{D0}$ in Section III.A).
 	Then given the dual variable in each iteration, we dedicatedly develop a two-stage method to determine $\bm{\alpha}$ and $\bm{\beta}$ for the dual problem, where the optimality will be theoretically proved in Section III.B.
 	Specifically, in the first stage, we subtly construct multiple subproblems (referring to $\mathbf{P1_{i,j}}$), each of which aims to independently seek the optimal KBC sub-policy (with respect to only $\alpha_{i}^{n}$ and $\alpha_{j}^{n}$) for each individual VUE pair.
 	After solving all these subproblems, an optimal coefficient matrix is obtained for all potential VUE pairs, by which we further construct a new subproblem (referring to $\mathbf{P2}$) in the second stage to find the optimal VSP strategy for $\bm{\beta}$.
  	
  	\subsection{Primal-Dual Problem Transformation}
    We first incorporate constraint (\ref{P0}e) into the objective function (\ref{P0}) by associating a Lagrange multiplier $\bm{\tau}=\{\tau_{i}|\tau_{i}\geqslant 0, i \in \mathcal{V}\}$, so that we have its associated Lagrange function as
    \begin{equation}
    	\begin{aligned}
    	L\left(\bm{\alpha},\bm{\beta},\bm{\tau}\right)& = \sum_{i \in \mathcal{V}}\sum_{j \in \mathcal{V}_{i}}\beta_{i\looparrowright j}\left(\delta_{i\looparrowright j}+\tau_{i}\theta_{i\looparrowright j}\right)-\theta_{0}\sum_{i \in \mathcal{V}}\tau_{i}\\
    	&\triangleq \widetilde{L}_{\bm{\tau}}(\bm{\alpha},\bm{\beta})-\theta_{0}\sum_{i \in \mathcal{V}}\tau_{i},~\label{Lagrangian} 
    	\end{aligned}
    \end{equation}
    where $\widetilde{L}_{\bm{\tau}}(\bm{\alpha},\bm{\beta})$ is defined for the sake of expression brevity.
    Then the Lagrange dual problem of $\mathbf{P0}$ is formulated by
     \begin{align}
			\mathbf{D0}:\ \max_{\bm{\tau}} \quad & D\left(\bm{\tau}\right)=g_{\bm{\alpha},\bm{\beta}}\left(\bm{\tau}\right)-\theta_{0}\sum_{i \in \mathcal{V}}\tau_{i},~\label{D}
	\end{align}
	where we have
	\begin{equation}
		\label{Dual}
			\begin{aligned}
			g_{\bm{\alpha},\bm{\beta}}\left(\bm{\tau}\right) \ &= \ \inf_{\bm{\alpha},\bm{\beta}} \ \widetilde{L}_{\bm{\tau}}(\bm{\alpha},\bm{\beta})\\
			{\rm s.t.} \ & \ \text{(\ref{P0}a)}-\text{(\ref{P0}d)}, \text{(\ref{P0}f)}, \text{(\ref{P0}g)}.
			\end{aligned}
	\end{equation}
	Notably, the optimality of the convex problem $\mathbf{D0}$ gives at least the best lower bound of $\mathbf{P0}$, even if $\mathbf{P0}$ is nonconvex, according to the duality property~\cite{boyd2004convex}.
	
	Given the initial dual variable $\bm{\tau}$, we can solve problem~(\ref{Dual}) in the first place to find the optimal solution $\bm{\alpha}$ and $\bm{\beta}$, the details of which will be presented later.
	After that, a subgradient method is employed in charge of updating $\bm{\tau}$ to solve $\mathbf{D0}$ in an iterative fashion, and generally, the convergence of subgradient descent can be ensured with the proper stepsize~\cite{boyd2003subgradient}.
	
	\subsection{Two-Stage Method Based on KBC and VSP}
    In this work, we develop a two-stage method to obtain the exactly optimal $\bm{\alpha}$ and $\bm{\beta}$ with a low computational complexity.
	In the first stage, we focus on multiple independent KB construction subproblems, each corresponding to a potential VUE pair in the SCVN.
	Specially, here the performances of the VUE $i$-VUE $j$ pair (i.e., the sender VUE $i$ and the receiver VUE $j$) and the VUE $j$-VUE $i$ pair (i.e., the sender VUE $j$ and the receiver VUE $i$) need to be considered together, and for ease of distinction, we refer to the two as a~\textit{VUE $i,j$ pair}, $\forall \left( i,j\right) \in \mathcal{V}\times \mathcal{V}_{i}, j>i$.
	That way, different KBC subproblems can be solved independently, so we let
	\begin{equation}
			\omega_{i,j}=\left(\delta_{i\looparrowright j}+\tau_{i}\theta_{i\looparrowright j}\right)+\left(\delta_{j\looparrowright i}+\tau_{j}\theta_{j\looparrowright i}\right).~\label{VUEpaircost}
	\end{equation}

	In this context, we now construct $U=\left(\sum_{i \in \mathcal{V}}\left|\mathcal{V}_{i}\right|\right)/2$ subproblems, each denoted as $\mathbf{P1_{i,j}}$ to seek the optimal KBC sub-policy only for an individual VUE $i,j$ pair.
	Herein, it is worth pointing out that the optimal KBC solution to problem~(\ref{Dual}) cannot be achieved by simply combining the obtained sub-policies of these $\mathbf{P1_{i,j}}$, but these sub-policies will be used to construct the subsequent VSP subproblem to finalize the joint optimal solution of $\bm{\alpha}$ and $\bm{\beta}$ for~(\ref{Dual}).
	Given the dual variable $\bm{\tau}$ in each iteration, $\mathbf{P1_{i,j}}$ becomes
	\begin{align}
		\mathbf{P1_{i,j}}:\ \min_{\left\{\alpha_{i}^{n}\right\},\left\{\alpha_{j}^{n}\right\}}\quad &\omega_{i,j}~\label{P1u}\\
		{\rm s.t.} \quad \quad \ & \sum_{n \in \mathcal{K}} \alpha_{i}^{n}\cdot s_{n}\leqslant C_{i},\tag{\ref{P1u}a}\\
		&\sum_{n \in \mathcal{K}} \alpha_{j}^{n}\cdot s_{n}\leqslant C_{j},\tag{\ref{P1u}b}\\
		&\eta_{i}\geqslant\eta_{0},\ \eta_{j}\geqslant\eta_{0},\tag{\ref{P1u}c}\\
		& \alpha_{i}^{n}\in \left\{ 0,1\right\}, \alpha_{j}^{n}\in \left\{ 0,1\right\}.\tag{\ref{P1u}d}
	\end{align}
	Since the problem scale of $\mathbf{P1_{i,j}}$ is quite small, here we directly employ a metaheuristic algorithm based on tabu search to quickly determine its solution~\cite{glover1989tabu}.
	Afterward, we can obtain the optimal KBC sub-policies for VUE $i$ (denoted as $\bm{\alpha}^{*}_{i_{(j)}}$) and VUE $j$ (denoted as $\bm{\alpha}^{*}_{j_{(i)}}$),\footnote{For auxiliary illustration, we use $(\cdot)$ in the subscript to specify the VUE pair attribute (relation) for each VUE's KBC sub-policy obtained from $\mathbf{P1_{i,j}}$.} corresponding to the individual VUE $i,j$ pair.
	The following proposition explicitly shows how the sub-policy of $\mathbf{P1_{i,j}}$ correlates to problem~(\ref{Dual}).
	
	\textit{Proposition 1:}
	Let $\bm{\alpha}^{*}=\left[\bm{\alpha}^{*}_{1}, \bm{\alpha}^{*}_{2}, \cdots,\bm{\alpha}^{*}_{V}\right]^{T}$ be the optimal KBC solution to the problem in~(\ref{Dual}) given the dual variable $\bm{\tau}$, where $\bm{\alpha}^{*}_{i}$ represents the optimal KBC policy of VUE $i$.
	Then we have $\forall i \in \mathcal{V}$, $\exists j \in \mathcal{V}_{i}$, such that $\bm{\alpha}^{*}_{i_{(j)}}=\bm{\alpha}^{*}_{i}$.
	\begin{IEEEproof}
			Given the optimal KBC solution $\bm{\alpha}^{*}$, let $\bm{\beta}^{*}=\left[\beta_{1\looparrowright j^{*}_{1}},\beta_{2\looparrowright j^{*}_{2}},\cdots,\beta_{V\looparrowright j^{*}_{V}}\right]^{T}$ be the corresponding optimal VSP solution to the problem in~(\ref{Dual}) under the same dual variable $\bm{\tau}$, where $\beta_{i\looparrowright j^{*}_{i}}$ ($\forall i \in \mathcal{V}$) indicates that VUE $j^{*}_{i}$ is the optimal SemCom node for VUE $i$, i.e., $\beta_{i\looparrowright j^{*}_{i}}=1$.
		
		From $\omega_{i,j}$ defined in~(\ref{VUEpaircost}), the objective function $\widetilde{L}_{\bm{\tau}}(\bm{\alpha},\bm{\beta})$ in~(\ref{Dual}) can be rewritten as $\widetilde{L}_{\bm{\tau}}(\bm{\alpha},\bm{\beta})=\sum_{i \in \mathcal{V}}\sum_{j \in \mathcal{V}_{i},j>i}\beta_{i\looparrowright j}\omega_{i,j}$, and then substituting $\bm{\beta}^{*}$ into $\widetilde{L}_{\bm{\tau}}(\bm{\alpha},\bm{\beta})$ can yield $\widetilde{L}_{\bm{\tau}}(\bm{\alpha},\bm{\beta}^{*})=\sum_{i \in \mathcal{V},i<j^{*}_{i}}\omega_{i,j^{*}_{i}}$, where $\omega_{i,j^{*}_{i}}$ is the term only related to VUE $i,j^{*}_{i}$ pair.
		
		Undoubtedly, if $\bm{\alpha}^{*}$ is further substituted into $\widetilde{L}_{\bm{\tau}}(\bm{\alpha},\bm{\beta}^{*})$, we can straightforwardly reach the optimality of the problem in~(\ref{Dual}).
		Since different VUE $i,j^{*}_{i}$ pairs are independent of each other, it means that different terms related to $\omega_{i,j^{*}_{i}}$ are independent of each other as well in $\widetilde{L}_{\bm{\tau}}(\bm{\alpha},\bm{\beta}^{*})$.
		Therefore, achieving the optimality of $\widetilde{L}_{\bm{\tau}}(\bm{\alpha},\bm{\beta}^{*})$ is equivalent to achieving the optimality of each $\omega_{i,j^{*}_{i}}$ under the given $\bm{\beta}^{*}$, where the optimality can be reached when $\bm{\alpha}=\bm{\alpha}^{*}$.
		
		In view of the above, we know that $\bm{\alpha}^{*}_{i}$ must be the optimal solution of $\omega_{i,j^{*}_{i}},\forall i \in \mathcal{V}$.
		Further combined with another fact that $\omega_{i,j}$ is the objective of $\mathbf{P1_{i,j}},\forall \left( i,j\right) \in \mathcal{V}\times \mathcal{V}_{i}, j>i$ where $\bm{\alpha}^{*}_{i_{(j)}}$ is the corresponding optimal solution, we can be sure that the equality $\bm{\alpha}^{*}_{i_{(j)}}=\bm{\alpha}^{*}_{i}$ holds when $j=j^{*}_{i}$.
	\end{IEEEproof}
	
	From Proposition 1, it is observed that the optimal KBC policy of each VUE can be found by solving a certain $\mathbf{P1_{i,j}}$.
	Hence, considering the single-association requirement, the optimal VSP strategy becomes the only key to finalize the optimal solution to~(\ref{Dual}).
	To achieve this, we first obtain the optimal coefficient matrix for $\bm{\beta}$ in~(\ref{Dual}) to account for all VSP possibilities.
	By calculating optimum $\omega_{i,j}$ (denoted as $\omega_{i,j}^{*}$) in $\mathbf{P1_{i,j}}$, the optimal coefficient matrix is formed as
	\begin{equation}
		\label{optimatrix}
		\bm{\Omega}=\left[
			\begin{matrix}
				+\infty &\omega_{1,2}^{*}&\omega_{1,3}^{*}&\cdots & \omega_{1,V}^{*}\\
				\omega_{2,1}^{*} &+\infty&\omega_{2,3}^{*}&\cdots & \omega_{2,V}^{*} \\
				\omega_{3,1}^{*} &\omega_{3,2}^{*}&+\infty&\cdots & \omega_{3,V}^{*} \\
				\vdots & \vdots & \vdots &\ddots & \vdots\\
				\omega_{V,1}^{*} & \omega_{V,2}^{*} & \omega_{V,3}^{*}& \cdots & +\infty
			\end{matrix}
			\right].
	\end{equation}
	$\bm{\Omega}$ is a $V\times V$ symmetric matrix where $\omega_{i,j}^{*}=\omega_{j,i}^{*}$, and all elements on its main diagonal are set to $+\infty$ to indicate the fact that a vehicle cannot communicate with itself, i.e., $j\neq i$.
	Besides, note that some $\omega_{i,j}^{*}$s in $\bm{\Omega}$ also have a value $+\infty$ if VUE $j$ is not the direct neighbor of VUE $i$, i.e., $j \notin \mathcal{V}_{i}$.
	
	Next, we concentrate upon the optimal vehicle service pairing strategy by constructing a new subproblem in the second stage.
	In line with the objective $\widetilde{L}_{\bm{\tau}}(\bm{\alpha},\bm{\beta})$ and $\bm{\beta}$-related constraints in~(\ref{Dual}), the VSP subproblem is written as
	\begin{align}
		\mathbf{P2}:\ \min_{\bm{\beta}}\quad &\frac{1}{2}\sum_{i \in \mathcal{V}}\sum_{j \in \mathcal{V}_{i}}\beta_{i\looparrowright j}\omega_{i,j}^{*}~\label{P2}\\
		{\rm s.t.} \quad & \text{(\ref{P0}c)},\text{(\ref{P0}d)}, \text{(\ref{P0}g)}.\tag{\ref{P2}a}
	\end{align}
	Given the dual variable, the optimal VSP strategy $\bm{\beta}$ (denoted as $\bm{\beta}^{*}=\left[\beta_{1\looparrowright j^{*}_{1}},\beta_{2\looparrowright j^{*}_{2}},\cdots,\beta_{V\looparrowright j^{*}_{V}}\right]^{T}$) can be directly finalized by solving $\mathbf{P2}$, where $\beta_{i\looparrowright j^{*}_{i}}$ ($\forall i \in \mathcal{V}$) indicates that VUE $j^{*}_{i}$ is the optimal SemCom node for VUE $i$, i.e., $\beta_{i\looparrowright j^{*}_{i}}=1$.
	To solve this problem, we first relax $\bm{\beta}$ into a continuous variable between $0$ and $1$ to make $\mathbf{P2}$ a linear problem, which can be efficiently solved with toolboxes such as CVXPY~\cite{diamond2016cvxpy}.
	Afterward, we restore the binary nature of the obtained continuous solution (denoted as $\bm{\beta}^{R}$) by
	\begin{equation}
		\label{VSPsolution}
		\beta_{i^{'}\looparrowright j^{'}}^{*}=\beta_{j^{'}\looparrowright i^{'}}^{*}=1,
	\end{equation}
	if and only if
	\begin{equation}
		\label{VSPsolution1}
		\left(i^{'},j^{'}\right)=\arg \max_{i\in \mathcal{V},j\in \mathcal{V}_{i},j>i}\beta_{i\looparrowright j}^{R}.
	\end{equation}
	Meanwhile, for the remaining $\beta_{i\looparrowright j}^{*}$ with respect to VUE $i^{'}$ and VUE $j^{'}$, we naturally have
	\begin{equation}
		\label{VSPsolution2}
			\left\{
			\begin{aligned}
			\beta_{i^{'}\looparrowright j}^{*}=\beta_{j\looparrowright i^{'}}^{*}=0, \quad & \forall j \in \mathcal{V}_{i^{'}},j \neq j^{'}\\
			\beta_{j^{'}\looparrowright i}^{*}=\beta_{i\looparrowright j^{'}}^{*}=0, \quad & \forall i \in \mathcal{V}_{j^{'}},i \neq i^{'}
			\end{aligned}.
		\right.
	\end{equation}
	Then we let $\mathcal{V}=\mathcal{V}\backslash \{i,j\}$, and repeat the above progresses until determining the optimal VSP solution for all VUEs.
	Finally, the obtained $\bm{\beta}^{*}$ is fed back to $\bm{\Omega}$ in order to further finalize the optimal KBC policy $\bm{\alpha}^{*}$.
	In the context of the solution to $\mathbf{P1_{i,j}}$, the approach to finalize $\bm{\alpha}^{*}$ can be stated more precisely as: for any $i \in \mathcal {V}$, we directly have $\bm{\alpha}_{i}^{*}=\bm{\alpha}^{*}_{i_{(j^{*}_{i})}}$.

	\section{Numerical Results and Discussions}
	In this section, numerical simulations are conducted to evaluate the performance of the proposed solution S$^{\text{4}}$ in SCVNs.
	We model a six-lane freeway (lane width: $4$ m) passing through a single cell with the RSU at its center (cell radius: $500$ m), where a total of $60$ VUEs are dropped according to the spatial Poisson process~\cite{3GPPLTE}.
	As for the settings relevant to SemCom, a total of $12$ different KBs are preset to provide VUEs with a variety of distinct services, and each of them has a storage size randomly distributed from $1$ to $5$ units.
	Correspondingly, we set a uniform KB storage capacity of $24$ units for all VUEs.
	Besides, each VUE's preference ranking for all KBs (i.e., $r_{i}^{n}$) is generated independently and randomly, where assuming that their respective Zipf distributions have the same skewness $1.0$.
	Likewise, either the average arrival rate of total semantic data packets, or the average interpretation time for packets based on the same KB $n$, is considered to be the same for all VUEs.
	Here, we fix the average total arrival rate $\lambda_{i}$ at $100$ packets/s and randomly generate the value of $1/\mu_{i}^{n}$ in a range of $5\times 10^{-3}\sim 1\times 10^{-2}$ s/packet with respect to different KB $n$.
	Further, the minimum knowledge preference satisfaction threshold $\eta_{0}$ and the maximum knowledge mismatch degree threshold $\theta_{0}$ are prescribed as $0.5$ and $0.1$, respectively.
	
	For comparison purposes, two different benchmarks are used in the simulations: 1) Distance-first pairing (DFP) strategy which assumes each VUE to choose its nearest unpaired VUE for V2V pairing; 2) Knowledge-first pairing (KFP) in which each VUE selects its neighboring unpaired VUE with the highest KB matching degree for V2V pairing.
	Among them, a personal preference-first KBC policy is considered for both benchmarks, which allows each VUE to construct KBs with the highest preferences until $\eta_{0}$ is satisfied, and then randomly select these unconstructed KBs until reaching respective maximum capacity.
	\begin{figure}[t]
		\centering
		\includegraphics[width=0.381\textwidth]{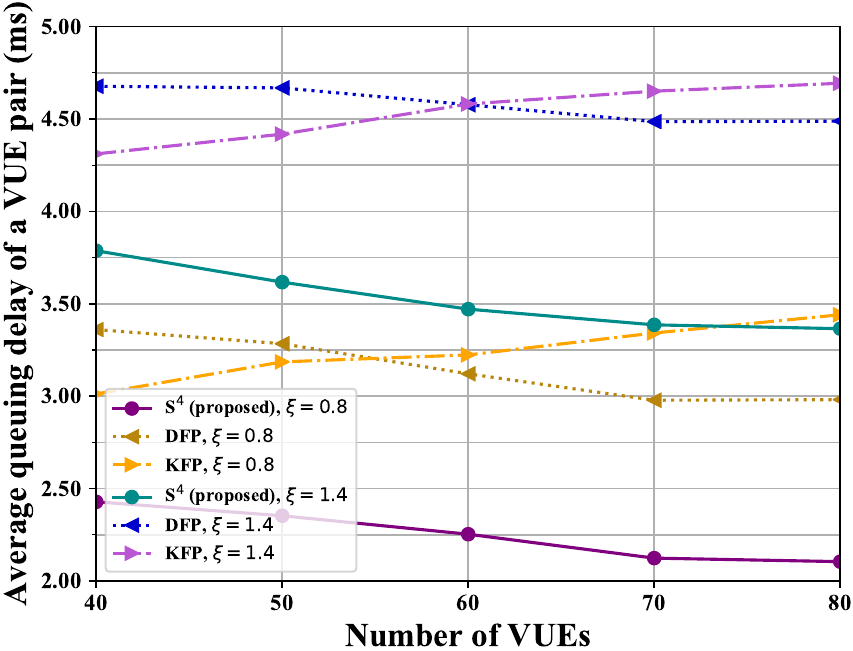} 
		\caption{Average queuing latency of a VUE pair vs. varying numbers of VUEs.}
		\label{SimFig1}
    \end{figure}

    Fig.~\ref{SimFig1} first depicts the average queuing latency performance of a VUE pair against varying numbers of VUEs, where two different KB preference skewness $\xi=0.8$ and $\xi=1.4$ are considered.
    In this figure, the latency of S$^{\text{4}}$ declines at the beginning with the number of VUEs, then remains stable beyond $70$ VUEs, and it can always outperform both benchmarks with an average latency reduction of around $1$ ms at any $\xi$. 
    The rationale behind this trend is that the more neighbors each VUE can have, the better chance of getting the low queuing latency for each VUE pair, which will be eventually stabilized when reaching the respective best achievable latency with a fixed bandwidth budget.
    Moreover, it is also observed that a larger $\xi$ causes a higher latency penalty, since the vast majority of VUEs' KBC is concentrated on a small number of KBs when $\xi$ increases.
    Clearly, a larger $\xi$ will make each participant more difficult to find a highly pairing VUE with the low latency under the given knowledge mismatching requirement $\theta_{0}$, thus resulting in a degraded performance.
    
    \begin{figure}[t]
		\centering
		\includegraphics[width=0.381\textwidth]{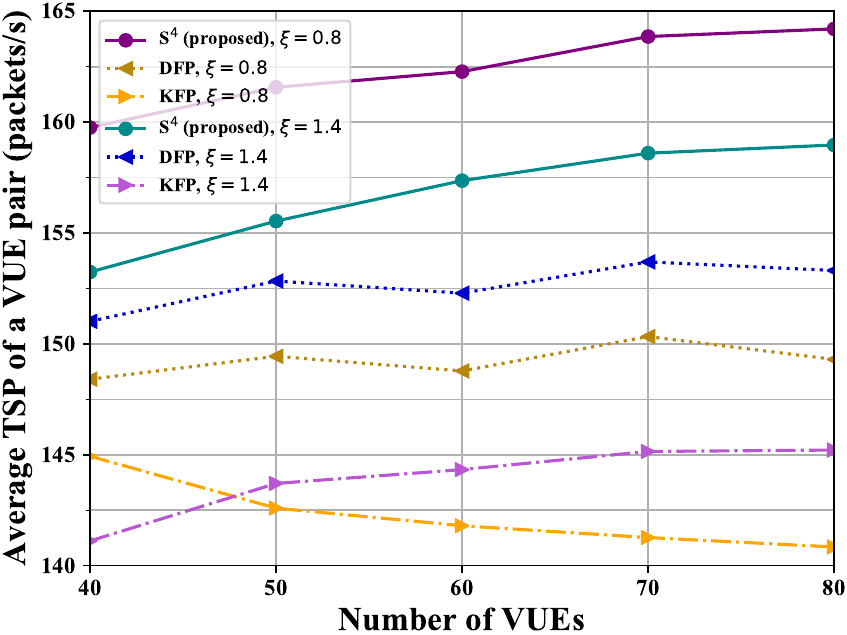} 
		\caption{Average TSP of a VUE pair vs. varying numbers of VUEs.}
		\label{SimFig2}
    \end{figure}
    The above analysis also applies to Fig.~\ref{SimFig2}, which compares all the three methods under the same settings as Fig.~\ref{SimFig1} to demonstrate the performance of the average throughput in semantic packets (TSP).
    Specifically, the TSP represents the total number of semantic packets that can be interpreted at a VUE pair per second, whose value is determined based on $\mathds{E}\left[W_{i\looparrowright j}\right]$ in~(\ref{PK1}).
    Likewise, the higher TSP is obtained as the number of VUEs increases, and our S$^{\text{4}}$ is still far better than the benchmarks at any point, e.g., with an average performance gain of $14$ packets/s compared with DFP and $20$ packets/s with KFP at $\xi=0.8$.
    Again, we see a better TSP when the KB popularity is diluted by a smaller $\xi$.
    \begin{figure}[t]
		\centering
		\includegraphics[width=0.381\textwidth]{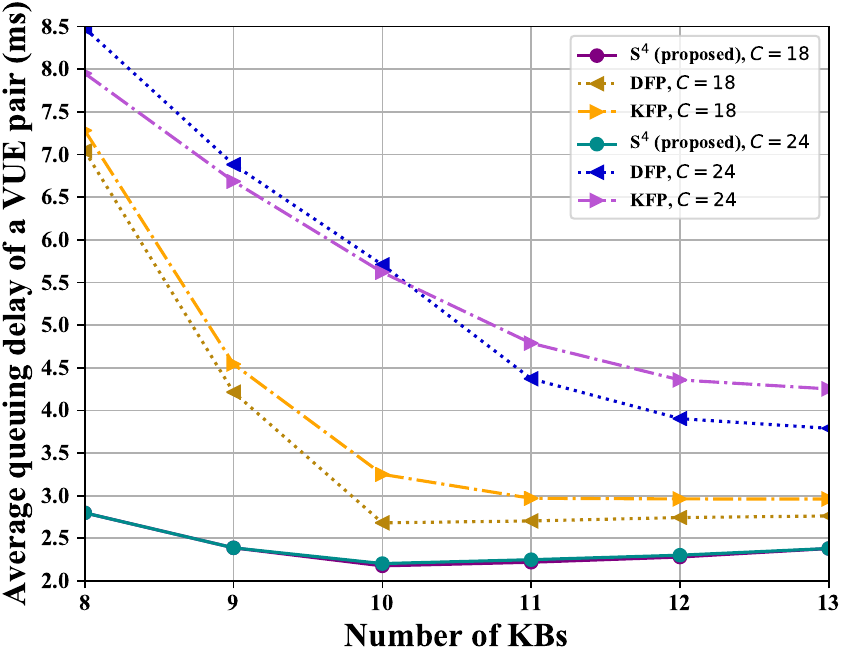} 
		\caption{Average queuing latency of a VUE pair vs. varying numbers of KBs.}
		\label{SimFig3}
    \end{figure}
    
    \begin{figure}[t]
		\centering
		\includegraphics[width=0.381\textwidth]{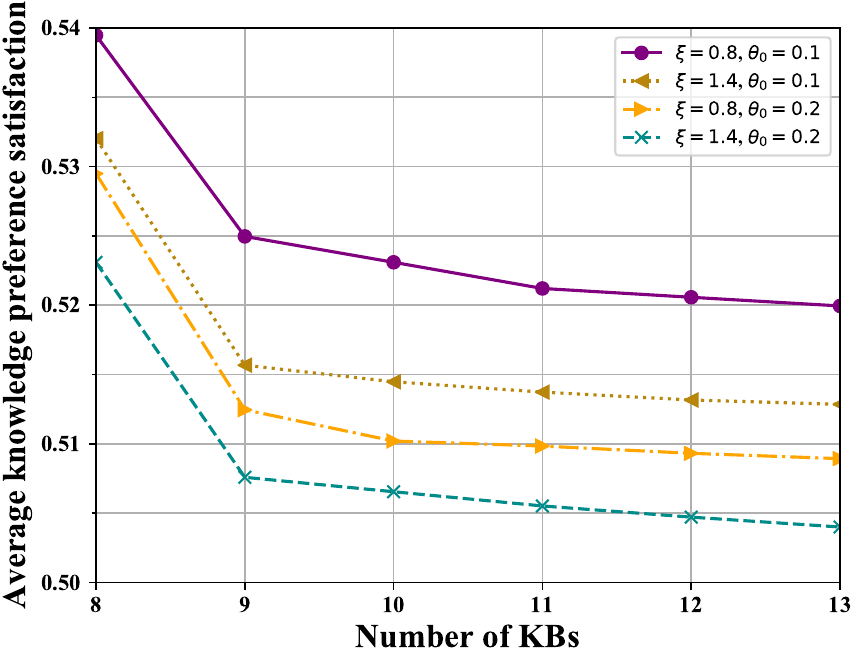} 
		\caption{Average knowledge preference satisfaction vs. varying numbers of KBs.}
		\label{SimFig5}
    \end{figure}
    Next, we explore the impact of varying number of KBs on the average queuing latency of a VUE pair with different VUE capacities $C=18$ and $C=24$, as demonstrated in Fig.~\ref{SimFig3}.
    It can be found that the latency drops fast at the beginning, and then rises slightly after exceeding $10$ KBs, whereas the performance of our S$^{\text{4}}$ still surpasses the benchmarks.
    This trend is attributed to the fact that more KBs imply less discrepancy in VUEs' preferences for different KBs given the fixed $\xi$, thereby at first leading to the higher probability for two paired VUEs constructing the KBs with high interpretation rates so as to render a lower delay.
    However, such performance gains will saturate and even worsen when these KBs with low interpretation rates become inevitably dominant in order to meet the minimum knowledge preference satisfaction threshold $\eta_{0}$.
    Besides, it is seen that different VUE capacities have little effect on the latency of S$^{\text{4}}$, although the larger capacity makes more KBs capable of construction.
	This is due to the latency-minimization objective we particularly focus on in the delay-sensitive SCVN, and only the KBs with low interpretation time should be selected.
    
    Finally, we validate the average knowledge preference satisfaction $\bar{\eta}=\frac{1}{V}\sum_{i \in \mathcal{V}}\eta_{i}$ reached at each VUE with varying numbers of KBs as shown in Fig.~\ref{SimFig5}, where $\xi = 0.8$, $\xi = 1.4$, $\theta_{0}=0.1$, and $\theta_{0}=0.2$ are taken into account.
    As the number of KBs increases, a lower $\bar{\eta}$ is obtained, which is to prevent these unnecessary KBs from being constructed while satisfying $\eta_{0}$ to the greatest extent.
    For the two curves with different $\xi$, referring to the analysis of Fig.~\ref{SimFig1}, a higher $\xi$ indicates a more concentrated KB preference, which means some extra KBs need to be constructed to meet the maximum $\theta_{0}$ requirement.
   	 Because of this, we also see a lower $\bar{\eta}$ at a higher $\theta_{0}$, since a more tolerable knowledge mismatch degree is more likely to avoid the unnecessary KBC.

	\section{Conclusions}
	In this paper, we proposed a novel solution S$^{\text{4}}$ to address the SemCom-empowered service provisioning problem in the SCVN.
	First, the KB matching based queuing latency expression of semantic data packets was derived, and then we identified and formulated the fundamental problem of KBC and VSP to minimize the queuing latency for all VUE pairs.
	After the primal-dual problem transformation, a two-stage method was developed specifically to solve multiple subproblems related to KBC and VSP with low computational complexity, and the solution optimality has been theoretically proved.
	Numerical results verified the sufficient performance superiority of S$^{\text{4}}$ in terms of both latency and reliability by comparing it with two different benchmarks.
	Accordingly, this work can be served as a preliminary research of applying SemCom to next-generation vehicular networks.

	\bibliographystyle{IEEEtran}
	\bibliography{main}

\begin{thebibliography}{10}

\bibitem{weaver1953recent}
W. Weaver, ``Recent contributions to the mathematical theory of communication,''
\newblock {\em ETC: A Review of General Semantics}, pp.~261--281, 1953.

\bibitem{bao2011towards}
J. Bao, P. Basu, M. Dean, C. Partridge, A. Swami, W. Leland, and J.~A. Hendler,
``Towards a theory of semantic communication,''
\newblock {in \em 2011 IEEE Network Science Workshop}, IEEE, 2011, pp.~110--117.

\bibitem{xia2022wiservr}
L. Xia, Y. Sun, C. Liang, D. Feng, R. Cheng, Y. Yang, and M. A. Imran, ``WiserVR: Semantic communication enabled wireless virtual reality delivery,''
\newblock {\em IEEE Wireless Communications}, vol.~30, no.~2, pp.~32--39, 2023.

\bibitem{strinati20216g}
E.~C. Strinati and S. Barbarossa,
``6G networks: Beyond Shannon towards semantic and goal-oriented communications,''
\newblock {\em Computer Networks}, vol.~190, p.~107930, 2021.

\bibitem{xia2022wireless}
L. Xia, Y. Sun, D. Niyato, X. Li, and M. A. Imran, ``Joint user association and bandwidth allocation in semantic communication networks,''
\newblock {\em IEEE Transactions on Vehicular Technology}, 2023.

\bibitem{9797984}
L. Xia, Y. Sun, X. Li, G. Feng, and M. A. Imran, ``Wireless resource management in intelligent semantic communication networks,''
\newblock {in \em IEEE INFOCOM 2022 - IEEE Conference on Computer Communications Workshops (INFOCOM WKSHPS)}, 2022, pp.~1--6. 

\bibitem{piantadosi2014zipf}
S. T. Piantadosi,
``Zipf's word frequency law in natural language: A critical review and future directions,''
\newblock {\em Psychonomic Bulletin \& Review}, vol.~21, no.~5, pp.~1112--1130, 2014.

\bibitem{lavee2009understanding}
G. Lavee, E. Rivlin, and M. Rudzsky,
``Understanding video events: A survey of methods for automatic interpretation of semantic occurrences in video,''
\newblock {\em IEEE Transactions on Systems, Man, and Cybernetics, Part C (Applications and Reviews)}, vol.~39, no.~5, pp.~489--504, 2009.

\bibitem{ross2014introduction}
S. M. Ross,
\newblock {\em Introduction to Probability Models}. Academic Press, 2014.

\bibitem{boyd2004convex}
S. Boyd, S. P. Boyd, and L. Vandenberghe,
\newblock {\em Convex Optimization}. Cambridge University Press, 2004.

\bibitem{boyd2003subgradient}
S. Boyd, L. Xiao, and A. Mutapcic,
``Subgradient methods,''
{\em Stanford University, Autumn Quarter}, vol.~2004, pp.~2004--2005, 2003.

\bibitem{glover1989tabu}
F. Glover,
``Tabu search-part I,''
{\em ORSA Journal on Computing}, vol.~1, no.~3, pp.~190--206, 1989.

\bibitem{diamond2016cvxpy}
S. Diamond and S. Boyd,
``CVXPY: A Python-embedded modeling language for convex optimization,''
{\em The Journal of Machine Learning Research}, vol.~17, no.~1, pp.~2909--2913, 2016.

\bibitem{3GPPLTE}
{\em 3GPP: Study on LTE-Based V2X Services; (Release 14)}, document TR 36.885, V2.0.0, 3GPP, Jun. 2016.

\end{thebibliography}
\end{document}